\documentclass{elsart}
\usepackage{epsfig}
\usepackage{supertabular}

\journal{Computer Networks}

\begin{document}
\begin{frontmatter}
\title {Internet Traffic Periodicities and Oscillations: A Brief Review}
\author{Reginald D. Smith}
\address {Bouchet-Franklin Research Institute, PO Box 10051, Rochester, NY 14610}
\date{April 23, 2009}
\ead {rsmith@bouchet-franklin.org}


\begin{abstract}
Internet traffic displays many persistent periodicities
(oscillations) on a large range of time scales. This paper describes
the measurement methodology to detect Internet traffic periodicities
and also describes the main periodicities in Internet traffic.
\end{abstract}

\end{frontmatter}

\maketitle

\section{Introduction}

Internet traffic has exploded in the last fifteen years as an area
of intense theoretical and experimental research. As the largest
engineered infrastructure and information system in human history,
the Internet's staggering size and complexity are reinforced by its
decentralized and self-organizing structure. Using packets of
encapsulated data and a commonly agreed protocol suite, the Internet
has far outgrown its origins as ARPANET whose traffic has demanded
new models and ways of thinking to understand and predict.

Amongst the earliest discoveries were the researches of Leland and
Wilson \cite{selfsimilar1} who identified the non-Poisson nature of
Internet traffic. This was followed by the seminal paper of Leland,
Taqqu, Willinger, and Wilson \cite{selfsimilar2} which proved that
Internet packet interarrival times are both self-similar and portray
long-range dependence. Though self-similarity is present at all time
scales, it is most well-defined when traffic is stationary, an
assumption that can only last a few hours at the most. The lack of
stationarity on long time scales is due to one of the most widely
known periodicities (or oscillations) in Internet traffic, the
diurnal cycle with 12 and 24 hour peaks.

Internet periodicities are not new and have been well-studied since
the earliest days of large-scale measurements of packet traffic,
however, they rarely receive primary attention in discussions of
traffic and are often mentioned only as an aside or a footnote.
Gradually, however, they are gaining more attention. This new area
of research has been dubbed \emph{network spectroscopy}
\cite{spectrum1} or \emph{Internet spectroscopy}. In this paper,
they will take front and center as the most important periodicities,
as well as the techniques to measure them, are described.

\section{Detection Methodologies}

Identifying periodicities in Internet traffic is, in general, not
markedly different from standard spectral analysis of any time
series. The same cautions apply with sampling rates and the Nyquist
theorem to determine the highest identifiable frequency as well as
to be aware of possible aliasing. In addition, the sampling period
is important due to the large ranges of magnitudes the periods of
Internet periodicities occupy.

The standard method is covered in \cite{spectrum5c,spectrum5b}. A
continuous time series is collected and binned with a sampling rate
$p$ where the number of packets arriving every $p$ interval seconds
are counted. Next, to remove the DC component of the signal, every
time step has the mean of the entire time series subtracted from it.
Next you calculate the autocovariance (ACVF) of the adjusted time
series. where for a time series of $N$ sampling periods (total
sampling time $pN$) the ACVF, $c$ at lag, $k$ is defined as

\begin{equation}
c(k) = \sum^{N-k-1}_{t=0}(X(t)-\overline{X})(X(t+k)-\overline{X})
\end{equation}

with a typical lag range chosen of $0 < k < N/2$. Finally, a Fourier
transform is taken of the ACVF with maximum lag $M$ and the
periodogram created from the absolute value (amplitude) of the
Fourier series

\begin{equation}
P(f) = \left|\sum^{M-1}_{k=0}c(k)e^{-i2\pi fk}\right|
\end{equation}

A resulting periodogram (see figure \ref{bgp}) has several typical
features. First, low frequency $1/f$ noise can be present, again
testifying to the self-similar nature of the traffic. This can
sometimes obscure low-frequency periodicities in the data. Second,
are any periodicities, their harmonics, and occasionally even small
peaks perhaps representing nonlinear mixing of a sort between 
two periodicities, often with periods of different orders of magnitude.

\subsection{Wavelet methods}

Given the nonstationary nature of Internet traffic and the frequent
presence of transients, methods based on the Fourier transform can
only given an incomplete view of the periodic dynamics of Internet
traffic. In particular, especially for rapidly changing
periodicities such as those caused by RTT of flows, periodicities
may only be temporary before shifting, disappearing, or being
displaced. Wavelet methods have been developed in great theoretical
and practical detail in the last several decades to allow for the
analysis of a signal's periodic nature on multiple times scales.
Wavelet techniques will not be covered here in detail though there
are many good references \cite{wavelet1,wavelet2,wavelet3,wavelet4}.
The continuous wavelet function on the signal $x(t)$, here an
Internet traffic trace, is given for a mother wavelet, $\psi$ with
$a$ representing a stretching coefficient (scale) and $b$ represents
a translation coefficient (time)

\begin{equation}
T(a,b) =
\frac{1}{\sqrt{a}}\int^{\infty}_{-\infty}x(t)\psi^*\left(\frac{t-b}{a}
\right)dt
\end{equation}

\begin{figure}[h]
    \centering
    \includegraphics[height=2.5in, width=2.5in]{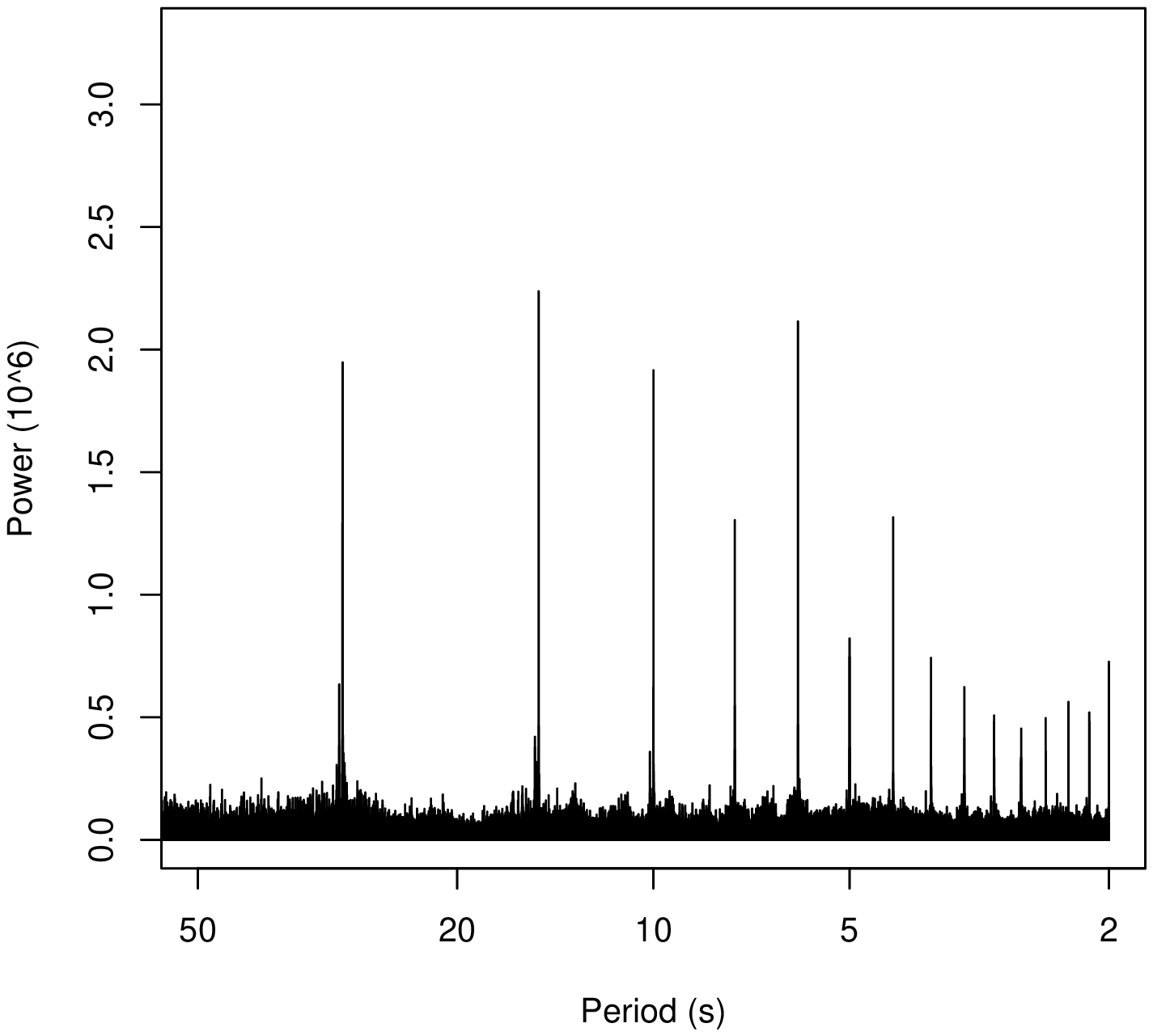}
    \includegraphics[height=2.5in, width=2.5in]{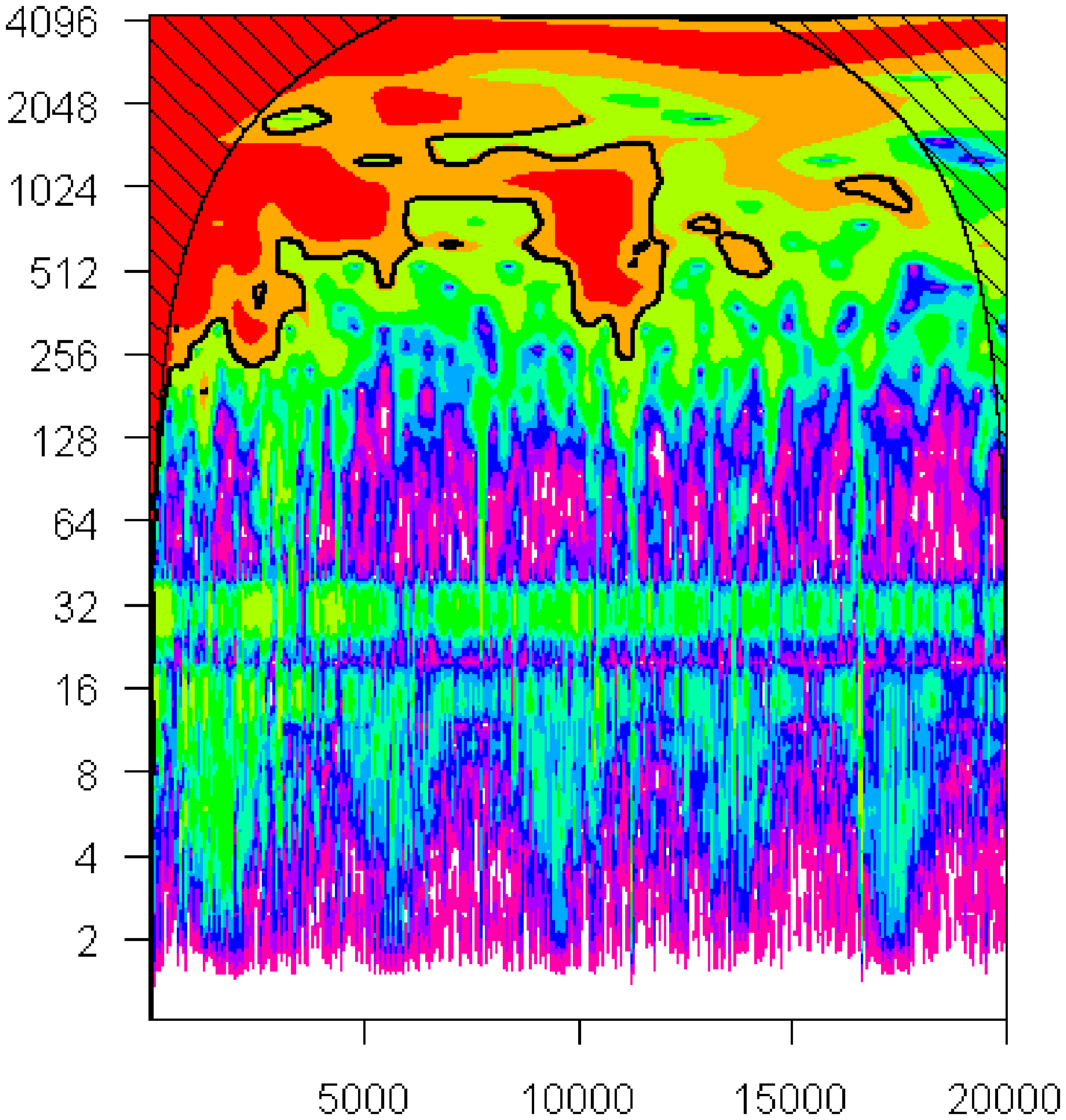}
        \includegraphics[height=2.5in, width=2.5in]{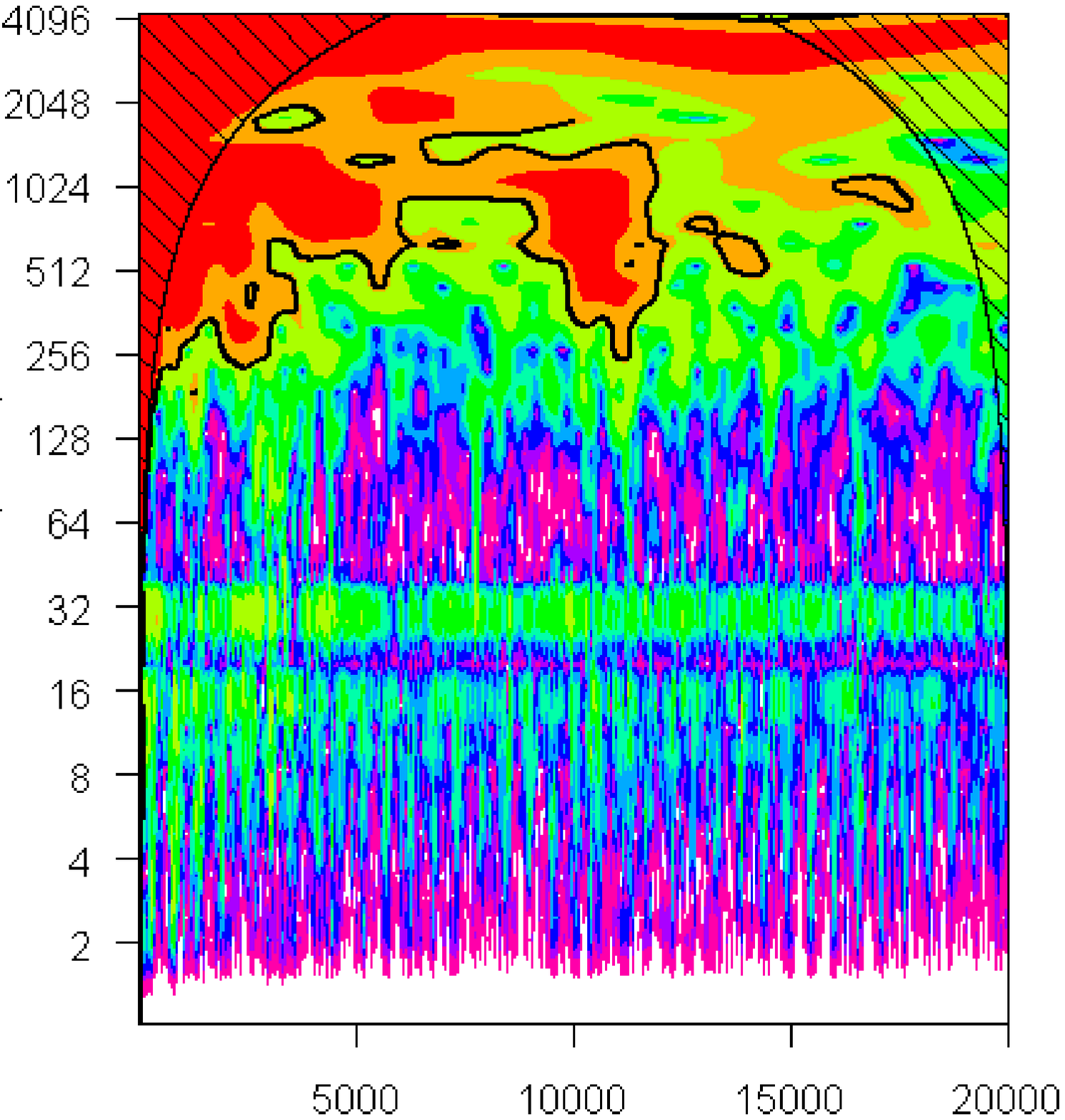}
        \caption{A FFT periodogram and two wavelet contour plots of the ACVF of BGP update packet traffic on node
        rrc00 on the RIPE Routing Information Service (RIS) BGP update traffic on February 1, 2009 \cite{ripe}.
        In the FFT plot, the strong peaks are due to the 30s BGP KEEPALIVE update packet messages with subsequent harmonics.
        The contour plot is based on a continuous wavelet transform using the Morlet wavelet for a 20,000s (5.5 hour) trace starting at 0000 GMT. The strong periodicity at 30s
        is evident, as well as the periodic bursts of high frequency traffic below it due to route flapping and the one hour
        periodic maximum suppression by route flap damping. The peaks correspond to the minimum penalty for the flapping route while
        the troughs correspond to the maximum penalty. In the third figure, the contour plot is recreated with the signal
        omitting packets that announce a withdrawal of one of the top 5 (most likely flapping) withdrawn IP addresses. The
        high frequency flapping is still present but not the coordinated hourly damping as in the second figure.}
    \label{bgp}
\end{figure}

In figure \ref{bgp} alongside the FFT of the signal is a contour
plot generated by plotting $T(a,b)$ using the Morlet mother wavelet
over 12 octaves. One of the key advantages of wavelets is seeing the
periodic variation over time. The y-axis represents the period of
the signal represented and the x-axis is the time of the traffic
trace in seconds.  A first feature is the continuous strong
periodicity at 30 seconds as a result of the update packets. A
second and more intriguing feature are the inverse triangular
`bursts' of high frequency traffic with an average period close to
one hour. These are update packets generated by route flapping,
which are damped for a maximum period of one hour according to the
most common presets for route flapping damping. The packets with the
most pernicious flapping routers announcing withdrawals were removed
in the third figure where the hourly oscillation largely disappears.

\section{Traffic oscillations/periodicities}

There are a plethora of traffic periodicities that represent
oscillations in traffic over periods of many orders of magnitude
from milliseconds to weeks. Broido, et. al. \cite{spectrum6} believe
there are thousands of periodic processes in the Internet. The sheer
range of the periods of the periodicities means that many times,
only certain periodicities appear in packet arrival time series due
either to the sampling rate or sampling duration. This is one of the
reasons why a comprehensive description of all Internet
periodicities has rarely been done.

Internet periodicities have origins which broadly correspond to two
general causes: first, there are protocol or data transmission
driven periodicities. These range on the time scale from
microseconds to seconds, or in rare cases, hours. These periodicties
can again be broken down into two smaller groups, periodicities
driven by packet data transmission on the link layer and
periodicities driven by protocol behavior on the transport layer.

Second are application driven periodicities. Their periods range on
the time scale of minutes to hours to weeks, and quite possibly
longer. These are all generated from activities at the application
layer, either by automated applications such as BGP or DNS or user
driven applications via HTTP or other user application protocols.

\begin{figure}[t]
    \centering
        \includegraphics[height=5in, width=7in]{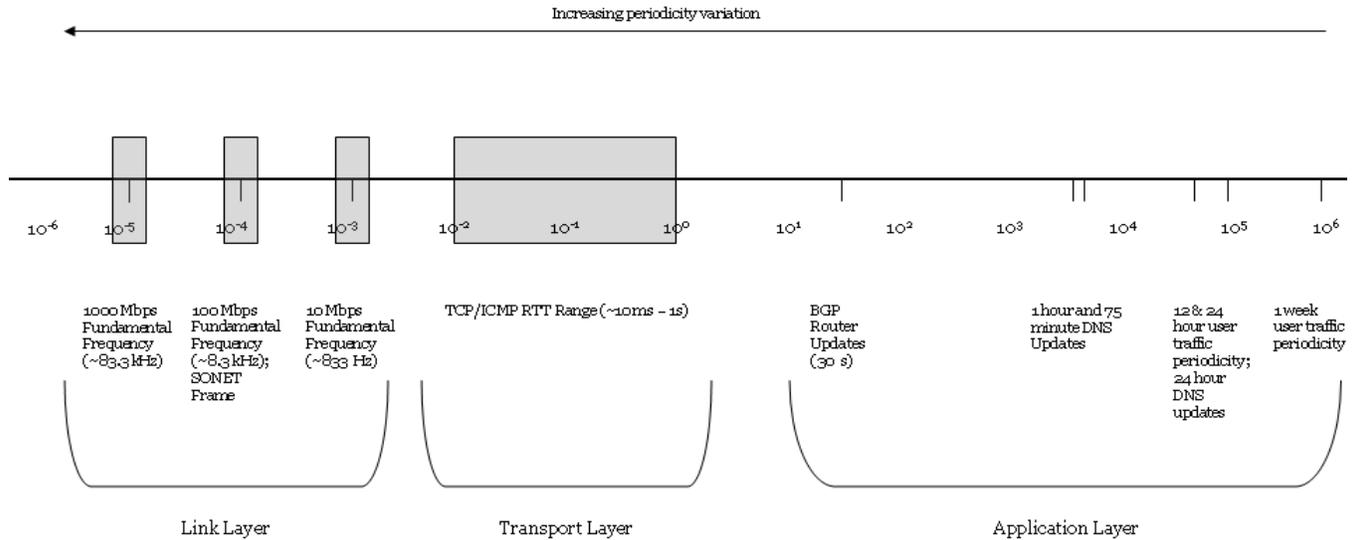}
        \caption{A rough breakdown of the major periodicities in Internet traffic showing the responsible protocols and
        their period in seconds. The periodicities span over 12 orders of magnitude and different protocol layers tend
        to operate on different time scales.}
    \label{periodicities}
\end{figure}

The major known periodicities are summarized in figure
\ref{periodicities} and will be described in detail in the next two
subsections.

\subsection{Key Periodicities: Link and Transport Layer}

A key link level periodicity due to the throughput of packet
transmission \cite{spectrum5, spectrum5b} of a link and can be
deduced from the equation:

\begin{equation}
\label{linkfreq} f = \frac{T}{s}
\end{equation}

Where $T$ is the average throughput of the link and $s$ is the
average packet size at the link level. The base frequency is the
rate of packet emission across the link at the optimum throughput
and packet size. The base frequency for data transmission is given
by

\begin{equation}
\label{linkfreqmax} f_{max}  = \frac{B}{MTU}
\end{equation}

where $B$ is the bandwidth of the link and the packet size is the
MTU packet size. Therefore for 1 Gigabit, 100 Mbps, and 10 Mbps
Ethernet links with MTU sizes of 1500 bytes, the theoretical optimal
base frequencies are 83.3 kHz, 8.3 kHz, and 833 Hz respectively.
Other technologies have their own specific periods such as SONET
frames identified with periods of 125$\mu$s\cite{spectrum1}.

These are among the most difficult traffic to identify due to the
need for high sampling rates of packet traffic. At a minimum, a
microsecond sampling rate is usually necessary to make sure you can
identify all link-layer periodicities. It is rare that both link
layer and other periodicities are displayed together since the
massive memory overhead of recording the timestamp of almost every
packet is necessary.

The link layer periodicities are receiving much of the attention in
the research, however, due to their possible use in inferring
bottlenecks and malicious traffic. The main practical applications
being researched are inferring network path characteristics such as
bandwidth, digital fingerprinting of link transmissions, and
detecting malicious attack traffic by changes in the frequency
domain of the transmission signal. \cite{spectrum10, spectrum10b}
use analysis of the distribution of packet interarrival times to
infer congestion and bottlenecks on network paths upstream. In
\cite{spectrum5b,spectrum5c, spectrum9, spectrum11,
spectrum12, spectrum14} various measures of packet
arrival distributions, particularly in the frequency domain, are
being tested to recognize and analyze distributed denial of service
or other malicious attacks against computer networks. Inspecting the
frequency domain of a signal can also reveal the fingerprints of the
various link level technologies used along the route of the signal
as is done in \cite{spectrum6, spectrum15}.

The transport layer also produces its own periodicities. In
particular, both TCP and ICMP often times operate bidirectional
flows with the interarrival of ACK packets corresponding to the RTT
between the source and destination \cite{spectrum6, spectrum9, spectrum8},
often in the range of 10 ms to 1 s. Instead of just
frequency peaks there usually are wide bands corresponding to the
dominant RTT in the TCP or ICMP traffic measured. According to most
equations of TCP throughput such as that by Semke et. al.
\cite{TCPthrough3} the throughput of TCP depends inversely on the
RTT so that the TCP RTT periodicities often can give a relative
estimate of throughput of the flows producing them and the
distribution of RTT for flows in the traffic trace. Exact estimates
are difficult though since packet loss and maximum segment size are
usually unknown. ICMP, though a connectionless protocol also has
echo replies which can also appear as periodicities if they are
persistent through time.

\subsection{Key Periodicities: Application Layer}

Once you rise to periods above one second, application layer
periodicities dominate the spectrum. These come from a variety of
sources including software settings and human activity. At the low
end are the 30s and sometimes 60s periodicities in BGP traffic. The
30s oscillation, shown in figure \ref{bgp}, is the most common set
time for routers to advertise their presence and continuing function
to neighboring routers using KEEPALIVE BGP updates. These are the
strongest periodicities present in BGP traffic. Large-scale
topological perturbations such as BGP storms can also produce
transient periodicities in traffic such as large-scale route
flapping which is shown in figure \ref{bgp}.

UDP traffic periodicities are rarely consistent and large-scale and
are generally generated by DNS, the largest application using UDP.
Claffy et. al. identified periodicities of DNS updates transmitted
with periods of 75 minutes, 1 hour, and 24 hours due to default
settings in Windows 2000 and XP DNS software\cite{spectrum1}. They
warn that such software settings could possibly cause problems in
Internet traffic if they lead to harmful periods of traffic
oscillations and congestion. Large numbers of usually source and
software specific UDP periodicities were also identified by Brondman
\cite{UDP}.

User traffic driven periodicities were the first known and most
easily recognized. The first discovered and most well-known
periodicity is the 24 hour diurnal cycle and its companion cycle of
12 hours. These cycles have been known for decades and reported as
early as 1980 and again in 1991 as well as in many subsequent
studies\cite{spectrum2,spectrum2b,spectrum2c,spectrum2d,spectrum2e,spectrum2f,spectrum3}.
This obviously refers to the 24 hour work-day and its 12-hour second
harmonic as well as activity from around the globe. The other major
periodicity from human behavior is the week with a period of 7 days
\cite{spectrum2c,spectrum2d,spectrum4} and a second harmonic at 3.5
days and barely perceptible third harmonic at 2.3 days. There are
reports as well of seasonal variations in traffic over months
\cite{spectrum5}, but mostly these have not been firmly
characterized. Long period oscillations have been linked to possible
causes of congestion and other network behavior related to network
monitoring \cite{spectrum2e, spectrum2f}. One note is that user
traffic driven periodicities tend to appear in protocols that are
directly used by most end users. The periodicities appear TCP/IP not
UDP/IP and are mainly attributable to activity with the HTTP and
SMTP protocols. They also often do not appear in networks with low
traffic or research aims such as the now defunct 6Bone IPv6 test
network.

\section{Discussion}

These periodicities range in roughly 12 orders of magnitude.
However, they share one particular characteristic. Namely, the
longer the period of the periodicity, the less likely it is to
betray variations in period or phase over time. For example, the
diurnal and weekly periodicities have their roots in human activity
and are based on the Earth's rotation and the seven week social
convention. These do not vary appreciably over long-time periods and
since they help drive human behavior which drives traffic, these
could be considered the most permanent of all periodicities and this
is partially why these were the earliest known. The BGP KEEPALIVE
updates and DNS updates are based on commonly agreed software
settings. These also do not vary appreciably and only change by user
preference. However, the transport and link layer periodicities are
much more variable. The RTT of TCP or ICMP varies depending on the
topological distance and congestion between two points. Hardly,
stable variables. Assuming the bandwidth of the link layers is
steady, the average packet size, which depends on both the maximum
transmission unit (MTU) software settings can cause large
variability to be seen in actual network traffic. Understanding the
range of these periodicities is more important than memorizing a
distance frequency value since it is always different depending on
the time and place of measurement. Internet periodicities will
likely play a large role in full characterization and simulation of
Internet traffic. Hopefully further work will put them in their
rightful place as fundamental phenomena of data traffic.

\end{document}